\documentclass[12pt]{iopart}
\usepackage{amsfonts}   
\usepackage{amssymb}

\begin{document}

\title[Tunneling Hamiltonian]{Tunneling Hamiltonian}

\author{J. Kern}

\address{Institut f{\"u}r Theoretische Physik, Universit{\"a}t Regensburg, 
93040 Regensburg, Germany}
\ead{johannes.kern@physik.uni-regensburg.de}
\begin{abstract}
For the description of the transport of electrons across a quantum dot, which 
is tunnel coupled to leads at different chemical potentials, it is usual to assume 
that the total Hamiltonian of the composite system of the leads and the quantum dot 
is the sum of three contributions: That of the leads (noninteracting electrons), that 
of the quantum dot and a third one, the ``tunneling Hamiltonian'', which reflects the 
possibility that electrons can move from the leads to the quantum dot or vice 
versa. The text aims at a mathematically clear derivation of such a separation.  
I will start the discussion with the total Hamiltonian of the 
system acting on a many-electron wave function, including the attractive interaction
between nuclei and electrons as well as the repulsive Coulomb-interaction between 
different electrons. Indeed, a natural separation of the total Hamiltonian
in the described form will be obtained. An analysis of the tunneling Hamiltonian 
shows that the electron-electron interaction yields contributions to it
which represent the correlated tunneling of {\em two} electrons at the same time.    
For the derivations it was useful to introduce a map called 
``antisymmetric product''. In an appendix I show possible exact representations 
of the total Hamiltonian (with arbitrary potential $V(r)$) obtained by the use 
of the antisymmetric product.   
\end{abstract}

\maketitle

\section{Hamilton operator}
The Hamilton operator of a crystal acts on an $N$-electron wave function 
\begin{equation}
 \psi : \quad D^N \quad \rightarrow \quad  \mathbb{C}, 
\end{equation} \label{wave function}
where $D = \mathbb{R}^3 \times \left\lbrace \uparrow, \downarrow 
\right\rbrace, D^N = D \times D \times \dots 
\times D$.
It contains the kinetic energy, the attractive force between the nuclei and the 
electrons and the repulsion 
between different electrons. It can be written as 
\begin{equation} \label{Hamiltonian}
 H \psi = - \frac{\hbar^2}{2m} \Delta \psi + \sum_{i = 1}^{N} V( r_i ) \psi 
+  \sum_{i < j} \frac{e^2}{\left| r_i - r_j \right|} \psi .
\end{equation}
The position of the nuclei is considered as constant in time within this equation 
\cite{Ashcroft}. I summarized in ``$V(r)$'' the potentials of the bare nuclei located 
at the lattice points of the crystal. 
A wave function as in (\ref{wave function}) can be viewed as a set of 
$2^N$ maps 
\begin{eqnarray*}
 \psi_{\sigma_1, \dots, \sigma_N} : \left( \mathbb{R}^3 \right)^N  & \rightarrow   
&  \mathbb{C}, \\
                        (r_1, \dots , r_N ) & \rightarrow & 
  \psi_{\sigma_1, \dots , \sigma_N } (r_1, \dots, r_N ).           
\end{eqnarray*}
The Hamiltonian leaves the spin-variables $\sigma_i \in \left\lbrace \uparrow, 
\downarrow \right\rbrace$ 
unchanged and acts on each of the $2^N$ components. 

The above shape of the Hamilton operator is quite general. It should be valid even 
if the nuclei are not 
located at the points of a regular lattice and also, for example, if one considers 
two finite crystals 
separated by vacuum. The changes enter into the potential $V(r)$. 

Lateron in the text it will turn out to be convenient to have an extra notation for 
the contribution of the electron-electron repulsion to the total Hamiltonian. I 
define thus 
\begin{equation} \label{interaction}
 H^{(int)} \psi = \sum_{i < j} \frac{e^2}{\left| r_i - r_j \right|} \psi . 
\end{equation}

\section{Antisymmetric wave functions}
\subsection{Slater determinants}
A wave function of $N$ fermions is required to be antisymmetric. Slater determinants 
\cite{Ashcroft} provide a way to construct antisymmetric wave functions of $N$ 
electrons by the use of one-electron wave functions: For a system of one-electron 
wave functions $\psi_i : D \rightarrow \mathbb{C}, i = 1, \dots, N,$ one can define
\begin{displaymath}
 \psi_1 \otimes \dots \otimes \psi_N : \quad D^N \rightarrow \mathbb{C}
\end{displaymath}
as 
\begin{displaymath}
 \psi_1 \otimes \dots \otimes \psi_N ( x_1, \dots , x_N ) := \frac{1}{\sqrt{N!}} 
\sum_{\pi \in S_N}
sgn (\pi) \psi_{\pi(1)} ( x_1 ) \dots \psi_{\pi (N)} ( x_N ). 
\end{displaymath}
I used ``$x_i$`` to denote elements of $D$, ''$S_N$'' is the set of all permutations 
of the set $\left\lbrace 1, \dots , N \right\rbrace$.

(For the set of permutations of any finite set the following rules 
hold: Any permutation can be represented as a product of subsequent 
``transpositions''. These are those special
permutations which exchange only two elements and act, apart from this, as identity. 
The representation of an arbitrary permutation as product of transpositions is not 
unique, but the property ``even`` or ``odd''of the number of needed transpositions 
is. Therefore, one can assign a sign to the permutation, $sgn (\pi) = \pm 1$, in 
case the number is even/odd. This sign is multiplicative, $sgn (\pi \pi')
= sgn ( \pi ) sgn ( \pi' )$. The ``Leibniz formula'' uses permutations to express 
the determinant. ) 

At any rate, the Slater determinant is an antisymmetric wave function. For any two
(quadratically integrable) wave functions $\psi, \psi' : D^N \quad \rightarrow 
\quad  \mathbb{C} $ one defines the scalar product \cite{Ashcroft} as
\begin{displaymath}
 \langle \psi | \psi' \rangle := \sum_{\sigma_1} \dots \sum_{\sigma_N} \int_{
\left( 
\mathbb{R}^3 \right)^N} 
\psi_{\sigma_1, \dots, \sigma_N}^* (r_1, \dots, r_N )
  \psi'_{\sigma_1, \dots , \sigma_N} (r_1, \dots ,r_N).
\end{displaymath}

\subsection{Normalization condition}
The norm is defined by $\parallel \psi \parallel^2 = \langle \psi | \psi  \rangle $. 
For simplicity, the prefactor $1/\sqrt{N!}$ was included in the Slater determinant. 
It ensures that, if the functions $\psi_1, \dots, \psi_N $ form an orthonormal 
system, then the norm of their Slater determinant is one. To fill the normalization 
condition with life, I tried to clarify its relation to the particle density. In 
the case of $N$ distinguishable particles represented by a wave function 
$\psi ( r_1, \sigma_1, \dots , r_N, \sigma_N )$ one could speak about a probability of 
finding particle $1$ with spin $\sigma_1$ within a volume $V_1 \subset \mathbb{R}^3$ 
and $\dots$ and particle $N$ with spin $\sigma_N$ within a volume 
$V_N \subset \mathbb{R}^3 $. The probability might be defined as 
\begin{displaymath}
 \int_{V_1}dr_1 \dots \int_{V_N} dr_N \quad \left| \psi ( r_1, \sigma_1, \dots , 
r_N, \sigma_N ) \right|^2.
\end{displaymath}
Correspondingly, one would define the number of particles which can be found in a 
volume $V \subset 
\mathbb{R}^3 $ as 
\begin{displaymath}
 \sum_{\sigma_1, \sigma_2} \left\lbrace \int_{V \times V^c} + \int_{V^c \times V } + 
2 \int_{V \times V}
\right\rbrace \left| \psi_{\sigma_1, \sigma_2} (r_1, r_2) \right|^2 .
\end{displaymath}
I assumed the special case $N=2$ and used the notation $V^c = \mathbb{R}^3 
\setminus V$ for this. One can then 
represent
\begin{displaymath}
 \int_{V \times V^c} + \int_{V^c \times V } + 2 \int_{V \times V} = 
\int_{V \times \mathbb{R}^3} + \int_{\mathbb{R}^3 \times V}.
\end{displaymath}
The way of defining the probability and representing it in an alternative form can 
be generalized to the case of arbitrary $N$. If the one-electron wave functions 
$\psi_1, \dots , \psi_N$ contributing to a Slater determinant 
form an orthonormal system, then one obtains indeed the particle density 
\begin{displaymath}
 \rho (r) = \sum_{i=1}^N  \sum_\sigma \left| \psi_i (r, \sigma)  \right|^2
\end{displaymath}
which one would expect intuitively.

\subsection{Antisymmetric product}
A Slater determinant can be viewed as an antisymmetric product of wave functions 
$\psi_i : D \rightarrow 
\mathbb{C}$. A generalization is possible: For 
\begin{displaymath}
 \psi : D^m \rightarrow \mathbb{C}, \quad \varphi: D^n \rightarrow \mathbb{C}
\end{displaymath}
one can define
\begin{displaymath}
 \psi \otimes \varphi : D^{m + n} \rightarrow \mathbb {C}
\end{displaymath}
by 
\begin{displaymath}
 \psi \otimes \varphi ( x_1, \dots , x_{m+n} ) := \sqrt{\frac{m!  n!}{(m+n)!}}
\sum_{M: M \subset \left\lbrace 1, \dots, m+n \right\rbrace, 
\left|M \right| = m}  sgn (\pi_M) 
\end{displaymath}
\begin{displaymath}
 \psi ( x_{\pi_M (1)}, \dots, x_{\pi_M (m)} ) \varphi ( x_{\pi_M (m+1)} , 
\dots, x_{\pi_M (m+n) } ). 
\end{displaymath}
I used ``$\left| M \right|$'' to denote the number of elements of $M$. The 
permutation $\pi_M$ is the one which counts at first the elements of $M$ in 
increasing order and then the elements of its complement 
$M^c := \left\lbrace 1, \dots , m+n \right\rbrace \setminus M $. More precisely: 
\begin{displaymath}
 \pi_M (1) = x_1, \dots, \pi_M (m) = x_m, \pi_M (m+1) = y_1, \dots , \pi_M (m+n) =  
y_n 
\end{displaymath}
if $M = \left\lbrace x_1, \dots, x_m \right\rbrace$,  $M^c = \left\lbrace y_1, 
\dots, y_n \right\rbrace $
in increasing order. 
By the property $sgn ( \pi_M ) sgn (\pi_{M^c} ) = (-1)^{mn}$ one can show the 
relation 
\begin{displaymath}
 \psi \otimes \varphi = (-1)^{m n} \varphi \otimes \psi.  
\end{displaymath}
If $\psi$ and $\varphi$ are antisymmetric, then also $\psi \otimes \varphi$. 
This can be seen by applying the ``antisymmetrizer'' \cite{Friedrich} 
to $\psi$ and $\varphi$.

The product is bilinear and associative: 
\begin{displaymath}
 \psi \otimes ( \varphi \otimes \lambda ) = ( \psi \otimes  \varphi ) \otimes \lambda .
\end{displaymath}
A symmetric representation is given by 
\begin{displaymath}
 \left[ \psi \otimes ( \varphi \otimes \lambda ) \right] (x_1, \dots , x_{m+n+l}) = 
\end{displaymath}
\begin{eqnarray*}
&=& \sqrt {\frac{m!n!l!}{(m+n+l)!}} \quad  
\sum_{M, N, L: \quad M \dot{\cup} N \dot{\cup} L =  \left\lbrace 1, \dots, 
m+n+l \right\rbrace  , |M| = m, |N| = n, |L| = l } sgn(\pi_{M,N,L}) \\
&& \psi ( x_{\pi_{M,N,L} (1)} , \dots , x_{\pi_{M,N,L} (m)} ) \\ &&   
  \varphi ( x_{\pi_{M,N,L} (m+1)} , \dots , x_{\pi_{M,N,L} (m+n)} ) \\ &&
\lambda ( x_{\pi_{M,N,L} (m+n+1)} , \dots , x_{\pi_{M,N,L} (m+n+l)} ), 
\end{eqnarray*}
where $\pi_{M,N,L}$ counts at first the elements of $M$, then the elements of 
$N$, and finally the elements of $L$ in increasing order.

As a consequence, expressions like ``$\psi_1 \otimes \dots \otimes \psi_N $'' 
(without any brackets) are
well-defined for $\psi_i : D^{m_i} \rightarrow \mathbb{C}$. Indeed, if all 
$m_i = 1$, then the product is 
the Slater determinant defined above.

\section{Vector space}
\subsection{Vector space of the leads}
About the vector space of the many-electron states of the leads I assume that it is 
generated by single-electron wave functions via Slater determinants and linear 
combinations of these. This is in agreement with the concept that the many-electron 
states in a large crystal can be specified by saying which ``one-electron levels'' 
are occupied. The Fermi-Dirac distribution was derived in Ref.  
\cite{Ashcroft} by the use of this idea. According to the Born-von Karman boundary 
condition one has two such one-electron wave functions for every ``allowed'' wave 
vector $k$ in a primitive cell of the reciprocal lattice. I denote the one-electron 
wave functions by 
\begin{eqnarray*}
 \psi_{l k \sigma} : \mathbb{R}^3 \times \left\lbrace \uparrow, \downarrow 
\right\rbrace 
& \rightarrow & \mathbb{C}  \\
(r, \sigma') & \mapsto &  \psi_{l k } (r) \delta_\sigma (\sigma').   
\end{eqnarray*}
The index ``$l$'' denotes the lead, $\psi_{lk} (r)$ is a Bloch wave function depending 
only on a space-variable. A further band index might be included but is not essential 
to the purpose of this work. The number of one-electron levels is proportional to 
the size of the crystal \cite{Ashcroft}. The $\psi_{lk\sigma}$ form an orthonormal 
system in the quadratically integrable functions $D \rightarrow \mathbb{C}$. For 
every natural number $n \le |R|$ one can define
\begin{displaymath}
 V_n (R) := lin \left\lbrace \psi_{1} \otimes \dots \otimes \psi_n : \psi_1, \dots, 
\psi_n \in R \right\rbrace. 
\end{displaymath}
(With ``$lin (S) $'' I denote the set of all linear combinations of elements in $S$.)
 $V_n (R)$ is a linear subspace of the set of all antisymmetric and quadratically 
integrable maps $D^n \rightarrow \mathbb{C}$. The subsets $M \subset R$ with 
$|M| = n$ form an orthonormal basis of $V_n(R)$ via
an identification
\begin{displaymath}
 M  =  \psi_1 \otimes \dots \otimes \psi_n, \mbox{ where } \psi_1, \dots, \psi_n 
\end{displaymath}
is a list of the elements of $M$ in some previously chosen and fixed order. The set 
of linear combinations of Slater determinants is really larger than the set of 
Slater determinants. In general, the solutions to the eigenvalue equation 
$H\psi = E \psi$ are linear combinations of an infinite number of Slater determinants. 
Any antisymmetric function can be approximated by a sequence of linear combinations 
of Slater determinants \cite{Friedrich}, while there is in general no way to 
approximate a linear 
combination of Slater determinants by a sequence of Slater determinants. 
The ground state energies of two-electron atoms have been calculated numerically 
by the use of the Hamilton operator, Eq. (\ref{Hamiltonian}) \cite{Friedrich}. 
The good quantitative agreement with experiments indicates that this rather 
intransparent Hamiltonian is indeed the right one as long as relativistic 
effects are ignored. Moreover, the Hartree-Fock equations have been derived from 
this Hamiltonian. They contain an {\em exchange term} which has been made responsible 
e.g. for screening effects.

Essentially in analogy to Ref. \cite{Bruus} I define the vector space $V(R)$ as 
\begin{displaymath}
 V (R) := V_0 (R) \times V_1 (R) \times \dots \times V_{|R|} (R),
\end{displaymath}
where $V_0 (R) := {\mathbb C}$. Since the $V_n(R)$ are Hilbert spaces via their 
scalar products $\langle . | . \rangle$, 
$V (R)$ is also a Hilbert space. $V_n(R)$ can be viewed as a subset of $V(R)$ by the 
identification
\begin{displaymath}
 \psi \in V_n(R)\quad \mbox{ corresponds to } \quad (0, \dots,0, \psi,0, \dots,0) 
\in V(R).  
\end{displaymath}
The element $(\psi_0, \dots, \psi_{|R|}) \in V(R)$ can be written as 
$\psi_0 + \dots + \psi_{|R|}$. The scalar
product is 
\begin{displaymath}
 \langle (\psi_0, \dots, \psi_{|R|} ) | (\varphi_0, \dots, \varphi_{|R|} ) \rangle 
= \sum_{i = 0}^{|R|} 
\langle \psi_i | \varphi_i \rangle.
\end{displaymath}
Moreover, one has the map
\begin{eqnarray*}
 \otimes: V(R) \times V(R) &\rightarrow& V(R) \\
 ( (\psi_0, \dots, \psi_{|R|} ) , (\varphi_0, \dots, \varphi_{|R|} ) )  &\mapsto& 
\sum_{i,j: \quad i + j \le |R| }  \psi_i \otimes \varphi_j.
\end{eqnarray*}
The elements of $V(R)$ can be viewed as linear combinations of subsets of $R$, in 
the same way as the elements of $V_n(R)$ can be viewed as such linear combinations. 
The element $1\in {\mathbb C} = V_0(R)$ represents the empty set. With this 
interpretation one can write for $M, N \subset R$:
\begin{displaymath} 
M \otimes N = \pm M\cup N \quad \mbox{ in case   } M \cap N = \emptyset    
\end{displaymath}
and zero otherwise.

\subsection{Approximate Hamiltonian}
The Hamiltonian ``$H^R$`` of the reservoirs is given by Eq. (\ref{Hamiltonian}) 
in case one replaces the potential $V(r)$ appearing there by the potential of the 
reservoirs ``$ V^R(r)$''. ($V(r) = V^R (r) + V^Q (r)$ with $V^Q(r)$ the potential 
which enters the Hamiltonian of the quantum dot.) There is little hope that $H^R$ is 
indeed an endomorphism of the Hilbert space $V(R)$, i.e., in general the image 
$H^R(V(R))$ is not a subset of $V(R)$. However, I assume that for states $\psi \in 
V(R) $ with ``sensible'' electron numbers $H^R \psi$ is very close to being an element 
of $V(R)$ and that in this sense $V(R)$ is a good choice of the vector space on which
the Hamiltonian operates. By the use of the orthogonal projection
\begin{displaymath}
 p_{V(R)} : \quad lin \left( V(R) \cup H^R (V(R)) \right) \rightarrow V(R),
\end{displaymath}
which can be defined by the condition
\begin{displaymath}
 \psi - p_{V(R)} \psi  \quad \bot \quad  V(R), 
\end{displaymath}
one obtains the approximate Hamiltonian $p_{V(R)} H^R$. This is indeed a hermitian 
operator $V(R) \rightarrow V(R)$ and defines the mathematical model of the reservoirs.

\subsection{Vector space of the composite system}
The construction of a vector space $V(M)$ can be realized for any finite set $M$
of one-electron wave functions. In case $M$ is an orthonormal system the subsets 
of $M$ form an orthonormal
basis of $V(M)$. I assume that the vector space of the many-electron states of the  
isolated quantum dot is given by $V(Q)$ where $Q$ is a finite orthonormal system of 
one-electron wave functions. Analogously to the above arguments, I assume that for 
the ``most relevant'' states $\varphi$ of the quantum dot $H^Q \varphi$ is very 
close to being an element in $V (Q)$. Hence, a good approximate Hamiltonian might be 
$p_{V(Q)} H^Q $.

I feel forced to assume that $R \cup Q$ is still an orthonormal system. This 
assumption is in agreement with previous tunneling theories \cite{Gottlieb}. Going 
even beyond this, I demand that 
\begin{eqnarray}  \label{orthogonality}
  H^R ( V(R)) \quad &\subset&  \quad \overline {V(Q^\bot)}, \nonumber \\ 
  H^Q ( V(Q)) \quad &\subset&  \quad \overline {V(R^\bot)}.
\end{eqnarray}
``$R^\bot$'' is the orthogonal complement of $R$ in the square-integrable functions 
$D \rightarrow \mathbb{C}$. $V(R^\bot)$ is the set of all linear combinations of 
Slater determinants of elements in $R^\bot$, in the same way as $V(R)$ is the set of 
all linear combinations of Slater determinants of elements in $R$. Finally, 
$\overline {V(R^\bot )}$ is the topological closure of $V(R^\bot) $, i.e., the set 
of all infinite and convergent linear combinations of Slater determinants of elements 
in $R^\bot$. The demand is that, in case one applies the Hamiltonian of the reservoirs 
to any wave function in the chosen vector space of the leads, then the result is a 
wave function for whose representation as an (infinite) linear combination of 
Slater determinants of one-electron wave functions exclusively elements in $ Q^\bot$ 
are needed, and vice versa. This should be approximately fulfilled if the distance 
between the sub-systems is large.

(Remark: $V(R)$ and $V(R^\bot)$ are not orthogonal since $\mathbb{C}$ is contained 
in both of them. Still, they are somehow antisymmetric complements since for all 
$\psi \in V(R), \varphi \in V(R^\bot): \parallel \psi \otimes \varphi \parallel^2 = 
\parallel \psi \parallel ^2 \parallel \varphi \parallel^2 $. One might take this 
equation as a definition of an antisymmetric complement ``$a.c.$`` and think about 
the question whether indeed $V(R)^{a.c.} = \overline {V(R^\bot)}$. )

As vector space of the many-electron states of the composite system of reservoirs 
and quantum dot $V(R \cup Q)$ should serve. In case of large distance between the 
two systems the assumption makes sense, for sure: The eigenstates of the composite 
system can be expected to be just the product of the eigenstates of the separate 
systems.

The Hamiltonian, Eq. (\ref{Hamiltonian}), acts on $V (R\cup Q)$ like
\begin{displaymath}
 H ( \psi_0, \psi_1, \dots , \psi_{|R\cup Q|} ) = ( 0, H \psi_1, \dots , 
H\psi_{|R\cup Q|} ),
\end{displaymath}
i.e., in particular $H\emptyset = 0$.   
As the final model Hamiltonian I take 
\begin{displaymath}
 p_{V(R\cup Q)} H: \quad V(R\cup Q) \rightarrow V (R \cup Q).
\end{displaymath}

The vector space 
$V(R\cup Q)$ is a realization of the ``tensor product`` of the vector spaces 
$V(R)$ and $V(Q)$, which is defined in mathematics in a purely formal way.   
The antisymmetric product is a realization of the corresponding bilinear map 
''$\otimes: V(R) \times V(Q) \rightarrow V(R\cup Q) $``. The purely formal tensor 
product has been used in Ref. \cite{Blum} to construct the vector space of a 
composite system of two sub-systems.

\section{Additive separation of the Hamiltonian}
For a start, I want to consider the action of the total Hamiltonian on an 
antisymmetric product $\psi \otimes \varphi \in V(R \cup Q)$ with $\psi \in V_m (R), 
\varphi \in V_n (Q),1 \le m \le |R|,1 \le n \le |Q|$. For this I write the 
electrostatic potential $V(r)$ caused by the nuclei entering the total Hamiltonian,
Eq. (\ref{Hamiltonian}), as $V(r) = V^Q (r) + V^R (r)$ where $V^S (r)$ is the 
electrostatic potential entering the Hamiltonian of the corresponding sub-system.  
The product $\psi \otimes \varphi$ can be written as
\begin{displaymath}
 \psi \otimes \varphi =  \sqrt {\frac{m!n!}{(m+n)!}} 
\sum_{M \subset \left\lbrace 1, \dots , m+n \right\rbrace: |M| = m} sgn(\pi_M)  
S_{\pi_M}
 \left[ \psi * \varphi \right] ,
\end{displaymath}
where $\psi * \varphi : D^{m+n} \rightarrow \mathbb{C}$ is defined by 
\begin{displaymath}
 \psi * \varphi (x_1, \dots, x_{m+n} ) = \psi ( x_1, \dots , x_m) \varphi 
( x_{m+1}, \dots, x_{m+n} )
\end{displaymath}
and where for any permutation $\pi \in S_{l},  l \in \mathbb {N},$ and any 
function 
\begin{displaymath}
 \alpha : D^{l} \rightarrow \mathbb {C}
\end{displaymath}
the function $S_\pi \alpha : D^l \rightarrow \mathbb{C}$ is given by 
\begin{displaymath}
 S_\pi \alpha ( x_1, \dots , x_l ) = \alpha ( x_{\pi (1)},  \dots, x_{\pi (l)} ).
\end{displaymath}
This representation is convenient since the Hamiltonian commutes with the 
operators $S_\pi$ 
\cite{Friedrich}, 
\begin{displaymath}
 H \left[ \psi \otimes \varphi  \right] = \sqrt {\frac{m!n!}{(m+n)!}} 
\sum_{M \subset \left\lbrace 1, \dots , m+n \right\rbrace: |M| = m} sgn(\pi_M)  
S_{\pi_M}
 H \left[ \psi * \varphi \right].
\end{displaymath}
With a little bookkeeping \cite{Ashcroft} one can write 
\begin{displaymath}
  H \left[ \psi \otimes \varphi  \right] = 
( H^R \psi ) \otimes \varphi + \psi \otimes (H^Q \varphi)
\end{displaymath}
\begin{equation} \label{tunneling_Hamiltonian}
 + 
\sqrt {\frac{m!n!}{(m+n)!}} 
\sum_{M \subset \left\lbrace 1, \dots , m+n \right\rbrace: |M| = m} sgn(\pi_M)  
S_{\pi_M}
 T_{m,n} \left[ \psi * \varphi \right],
\end{equation}
where $T_{m,n}$ is defined by 
\begin{displaymath}
 T_{m,n} := T_{m,n}^{(Q)} + T_{m,n}^{(R)} + T_{m,n}^{(int)}, 
\end{displaymath}
\begin{eqnarray*}
 T_{m,n}^{(int)} \left[ \alpha (r_1, \sigma_1, \dots , r_{m+n}, \sigma_{m+n} ) 
\right] := &&
\end{eqnarray*}
\begin{displaymath}
\left\lbrace   \sum_{i = 1}^m \sum_{j =m+1}^{m+n} 
\frac{e^2}{ |r_i - r_j| }  \right\rbrace
\alpha (r_1, \sigma_1, \dots , r_{m+n}, \sigma_{m+n} ), 
\end{displaymath}
\begin{eqnarray*}
 T_{m,n}^{(Q)} \left[ \alpha (r_1, \sigma_1, \dots , r_{m+n}, \sigma_{m+n} ) 
\right] := &&
\end{eqnarray*}
\begin{displaymath}
\left\lbrace  \sum_{i = 1}^m  V^Q (r_i)    \right\rbrace
\alpha (r_1, \sigma_1, \dots , r_{m+n}, \sigma_{m+n} ), 
\end{displaymath}
and where $T_{m,n}^{(R)}$ has the corresponding definition.

For the above alternative representation of  $H \left[ \psi \otimes 
\varphi  \right]$ the assumption $m, n \ge 1$ was made. The equality, however, 
is correct also in the cases that $m$ or $n$ or both of them are zero. (For example,
 $\psi \otimes \varphi = \psi \varphi $ in case $\psi$ or $\varphi $ is only a number; 
$T_{0,0 } =0$.) Roughly speaking, the terms in $T_{m,n}^{(int)}$ reflect the 
repulsion between electrons in the contacts on the one hand and electrons in the 
quantum dot on the other hand. The terms in $T_{m,n}^{(Q)} $ represent 
the attractive interaction of the electrons in the reservoirs on the one hand and 
the nuclei in the quantum dot on the other hand.

Since $p_{V(R\cup Q)} H$ was chosen as the model Hamiltonian, the projection 
$p_{V(R\cup Q)}$ should be applied to both sides of Eq. (\ref{tunneling_Hamiltonian}). 
By the demanded  complementarity (\ref{orthogonality}), one obtains the equation
\begin{displaymath}
 p_{V(R\cup Q)} H \left[ \psi \otimes \varphi  \right] = 
(p_{V(R)} H^R \psi ) \otimes \varphi + \psi \otimes (p_{V(Q)}H^Q \varphi)
\end{displaymath}
\begin{displaymath} 
 +  p_{ V (R \cup Q) } \quad
\sqrt {\frac{m!n!}{(m+n)!}} 
\sum_{M \subset \left\lbrace 1, \dots , m+n \right\rbrace: |M| = m} sgn(\pi_M)  
S_{\pi_M}
 T_{m,n} \left[ \psi * \varphi \right].
\end{displaymath}
\underline{Proof:} \newline
I used that $p_{V(R\cup Q)}   \left[ ( H^R \psi )  \otimes \varphi \right] 
=   ( p_{V(R)}  H^R \psi )  \otimes \varphi  $ for any $\psi \in V_m(R), \varphi 
\in V_n(Q), m, n \ge 0$. Thus, I ought to show that 
\begin{displaymath}
  ( H^R \psi   -  p_{V(R)}  H^R \psi )  \otimes \varphi \quad \bot 
\quad V(R\cup Q).  
\end{displaymath}
According to the complementarity-condition (\ref{orthogonality}),  $H^R \psi$ is a 
(maybe infinite) linear combination of Slater determinants of functions in $Q^\bot$. 
Since $R$ is a finite subset of $Q^\bot$, one can assume that all of the 
Slater determinants appearing in the expansion have the shape 
\begin{displaymath}
 \psi_1 \otimes \dots \psi_{i_0} \otimes \psi_{i_0 + 1} \otimes \dots \otimes 
\psi_{m},
\end{displaymath}
 where $ \psi_i \in R$ if $i \le i_0$, $\psi_i \bot R$ otherwise. If one applies 
the projection $p_{V(R)}$ to such a Slater determinant, then one obtains zero in case
$i_0 < m$. The Slater determinant remains unchanged in case $i_0 = m$. Thus,  
$H^R \psi   -  p_{V(R)}  H^R \psi$ is a linear combination of such determinants 
with $i_0 < m$. For any $\varphi \in V_n(Q)$ and any such Slater determinant, however,
one gets 
\begin{displaymath}
 \psi_1 \otimes \dots \psi_{i_0} \otimes \psi_{i_0 + 1} \otimes \dots \otimes 
\psi_{m} \otimes \varphi \quad \bot \quad V(R \cup Q).  \quad \Box
\end{displaymath}

For any two endomorphisms $A \in End(V(R)), B \in End(V(Q))$ ($End(V)$ is the set of 
linear maps $V \rightarrow V$) there is a unique well-defined endomorphism 
''$A\otimes B $`` $\in End( V(R \cup Q))$ which has the property:
\begin{displaymath}
A\otimes B ( \psi \otimes \varphi ) = (A\psi) \otimes (B \varphi) \mbox{ for all }
\psi \in V(R), \varphi 
\in V(Q).
\end{displaymath}
The proof of this statement is straightforward and perfectly analogous to the 
proof of the corresponding statement about formal tensor products. One can 
define:
\begin{eqnarray*}
H_{tot}      &:=& p_{V(R\cup Q)} H, \\
H_R    &:=& (p_{V(R)} H^R) \otimes 1_{V(Q)}, \\
H_Q    &:=&  1_{V(R)} \otimes (p_{V(Q)} H^Q),  
\end{eqnarray*}
and finally 
\begin{equation} \label{definition_of_H_T}
 H_T := H_{tot} - H_R - H_Q.
\end{equation}
All of these are hermitian operators on $V(R\cup Q)$.

\section{Creation- and annihilation operators}
By the use of creation- and annihilation
operators \cite{Bruus} $H_T $ can be expressed in a compact form. For any normalized
 element $\psi \in V_1 (R \cup Q)$ one can define the creation operator of $\psi$ 
as the map 
\begin{eqnarray*}
 c_\psi^\dagger : V(R \cup Q)  & \rightarrow & V(R \cup Q), \\
  \varphi &\mapsto & \psi \otimes \varphi .  
\end{eqnarray*}
The annihilation operator of the one-electron level $\psi$ is defined as the 
adjoint operator of the creation operator, $c_\psi := 
\left( c_\psi ^\dagger \right)^\dagger $.

\section{Expression for $H_T$ in terms of creation- and annihilation operators}
So far, a representation of the total Hamiltonian 
\begin{displaymath}
 H_{tot} = H_R + H_Q + H_T
\end{displaymath}
was obtained, where $H_R$ can be called the contribution of the leads, $H_Q$ is the 
contribution of the quantum dot and $H_T$ is the tunneling Hamiltonian. The main part 
of this text aims at a description of $H_T$. $H_T$ leaves the particle number 
unchanged, since the 
other contributions as well as the total Hamiltonian are doing so; 
$H_T \emptyset = 0$.

For the description of the linear operator $H_T$ one can use any basis of 
$V(R \cup Q)$. Independently of whether $H_R$ and $H_Q$ are diagonal in the 
Slater determinants given by the subsets of $R \cup Q$ one can thus use these 
Slater determinants for the description. For example, if $x_1, \dots , x_m $ 
are $m$ different elements in $R$, if $ y_1, \dots , y_n$ are $n$ different 
elements in $Q$  and if $z_1, \dots , z_{m+n} \in R\cup Q$ are $m+n$ different 
elements, then one obtains according to Eq. (\ref{tunneling_Hamiltonian})
\begin{displaymath}
 \langle z_1 \otimes \dots \otimes z_{m+n} | 
H_T \left[ x_1 \otimes \dots \otimes x_m \otimes y_1 \otimes \dots \otimes y_n 
\right]  \rangle  \quad = 
\end{displaymath}
\begin{eqnarray} \label{long_line_1}
 && \frac{1}{(m+n)!} \sum_{M \subset \left\lbrace 1, \dots, m+n \right\rbrace: 
|M| = m} 
  \quad \sum_{\pi \in S_{m+n}} \quad \sum_{\tau \in S_m} \sum_{\eta \in S_n} 
\nonumber \\  && 
sgn(\pi_M) sgn(\pi) sgn(\tau) sgn (\eta) \nonumber \\ &&
\langle z_{\pi(1)} * \dots * z_{\pi (m+n)} | \nonumber \\ &&
S_{\pi_M} T_{m,n} \left[ x_{\tau (1)} * \dots * x_{\tau (m)} *  
y_{\eta (1)} * \dots * y_{\eta (n)} \right] \rangle. 
\end{eqnarray}

\subsection{Contribution of the electron-electron interaction to $H_T$ } 
First, the contribution of $T_{m,n}^{(int)}$ to the matrix 
element is considered. Hence, I assume that $m, n \ge 1$. For any $1 \le i \le m, 
1 \le j \le n$, the permutation $k_{ij} \in S_{m+n}$ is defined by 
(myself and) the conditions:
\begin{itemize}
 \item $\kappa_{ij}(1) = i, \kappa_{ij}(2) = j$.
\item $\kappa_{ij} (3), \dots, \kappa_{ij}(m+n)$ is an increasing list of the 
remaining numbers.
\end{itemize}
For any sequence of one-electron wave functions one can write
\begin{displaymath}
 \frac{e^2}{|r_i - r_j|} v_1 * \dots * v_{m+n} = 
\end{displaymath}
 \begin{displaymath}
  S_{\kappa_{ij}} \left\lbrace \left[ H^{(int)} \left( 
v_{\kappa_{ij} (1)} * v_{\kappa_{ij} (2)} \right) 
\right]  * v_{\kappa_{ij} (3)} * \dots * v_{\kappa_{ij} (m+n)} \right\rbrace . 
 \end{displaymath}
For this, the operator of the electron-electron interaction defined in 
Eq. (\ref{interaction}) was used. Moreover, I define now
\begin{equation} \label{definition_of_w-s}
 w_1 := x_1, \dots , w_m := x_m, w_{m+1}:= y_1, \dots , w_{m+n} := y_n, 
\end{equation}
moreover the permutation ''$\tau \cdot \eta $`` $\in S_{m+n} $ by 
\begin{itemize}
 \item $(\tau \cdot \eta) (l) := \tau (l), l \le m$.
\item $ (\tau \cdot \eta) (l) := \eta (l-m) + m, l \ge m +1$.   
\end{itemize}

Inserting all of this into Eq. (\ref{long_line_1}), the contribution 
\begin{eqnarray} \label{long_line_2}
 && \frac{1}{(m+n)!} \sum_{M \subset \left\lbrace 1, \dots, m+n \right\rbrace: 
|M| = m} 
  \quad \sum_{\pi \in S_{m+n}} \quad \sum_{\tau \in S_m} \sum_{\eta \in S_n} 
\quad \sum_{i \le m}
\sum_{j \ge m+1}
\nonumber \\  && 
sgn(\pi_M) sgn(\pi) sgn(\tau) sgn (\eta) \nonumber \\ &&
\langle z_{\pi \pi_M (\tau \cdot \eta)^{-1} (\tau \cdot \eta ) 
\kappa_{ij}(1)} * \dots * z_{\pi \pi_M (\tau \cdot \eta)^{-1} (\tau \cdot \eta ) 
\kappa_{ij}(m+n) } 
| \nonumber \\ && 
 H^{(int)} \left[ w_{(\tau \cdot \eta) \kappa_{ij} (1)} * 
 w_{(\tau \cdot \eta) \kappa_{ij} (2)} \right] 
* 
w_{(\tau \cdot \eta) \kappa_{ij} (3)} 
\dots * w_{(\tau \cdot \eta) \kappa_{ij} (m+n)} \rangle 
\end{eqnarray}
of the electron-electron interaction to the matrix element is obtained. The fact 
that the operators $S_{\pi'}$ conserve the scalar product, $\langle S_{\pi'} \psi 
| S_{\pi'} \varphi \rangle = \langle \psi | \varphi \rangle$, and the relation
$S_{\pi'} S_{\pi''} = S_{\pi'  \pi''}$ were used for this. (In particular: 
$S_{\pi'}^{-1}  = S_{\pi'^{-1}} $.)

For all $1 \le i_0 \le m$,  $m+1 \le j_0 \le m+n $ and $  f_0, g_0  
\in \left\lbrace 1, \dots , m+n \right\rbrace, f_0 \neq g_0,$ 
I define the set $\Sigma_{i_0j_0}^{f_0 g_0}$ as the set of all tuples 
\begin{displaymath}
 (M, \pi, \tau, \eta, (i,j))
\end{displaymath}
 with the property: 
\begin{eqnarray*}
 (\tau \cdot \eta) \kappa_{ij} (1) &=& i_0, \\
 (\tau \cdot \eta) \kappa_{ij} (2) &=& j_0,  \\
 \pi \pi_M (\tau \cdot \eta)^{-1} (i_0) &=& f_0,  \\
 \pi \pi_M (\tau \cdot \eta)^{-1} (j_0) &=& g_0.
\end{eqnarray*}
The set of all tuples $(M, \pi, \tau, \eta, (i,j))$ over which the sum goes is 
the disjoint union of the sets $\Sigma_{i_0j_0}^{f_0 g_0}$. The sum 
(\ref{long_line_2}) turns into 
\begin{eqnarray} \label{long_line_3}
 &&  \sum_{(i_0, j_0, f_0, g_0)} \quad
\sum_{(M, \pi, \tau, \eta, (i,j)) \in  \Sigma_{i_0j_0}^{f_0 g_0}} 
 \frac { sgn(\pi \pi_M (\tau \cdot \eta)^{-1}) }{(m+n)!} \nonumber \\ &&
\langle z_{f_0} * z_{g_0} |H^{(int)} \left[ w_{i_0} * w_{j_0} \right] 
\rangle \nonumber \\ &&
\prod_{l \neq i_0, j_0} 
\langle z_{\pi \pi_M (\tau \cdot \eta )^{-1} (l)} | w_l \rangle .
   \end{eqnarray}

There is at most one permutation ``$\pi_{i_0j_0}^{f_0 g_0}$'' $\in S_{m+n}$
with the property that the product
\begin{equation} \label{properties}
 \prod_{l \neq i_0, j_0} 
\langle z_{ \pi_{i_0 j_0}^{f_0 g_0} (l)} | w_l \rangle  
\end{equation}
is nonzero and with the property that 
\begin{displaymath}
 \pi_{i_0 j_0}^{f_0 g_0} (i_0) = f_0, \quad \pi_{i_0 j_0}^{f_0 g_0} (j_0) = g_0.
\end{displaymath}
The value of the product is one in this case. {\em If} there is a permutation
$ \pi_{i_0 j_0}^{f_0 g_0} $ with the demanded properties, then the inner sum 
of expression (\ref{long_line_3}) turns into 
\begin{displaymath}
 sgn( \pi_{i_0,j_0}^{f_0, g_0} ) \quad  
\langle z_{f_0} * z_{g_0} |H^{(int)} \left[ w_{i_0} * w_{j_0} \right] 
\rangle ,
\end{displaymath}
since the number of tuples $(M, \pi, \tau, \eta, (i,j))$ in 
$ \Sigma_{i_0j_0}^{f_0 g_0} $ with the property  $\pi \pi_M 
(\tau \cdot \eta )^{-1} = \pi_{i_0 j_0}^{f_0 g_0} $ is then $(m+n)!$. Swapping 
the position of $ f_0 $ and $g_0$ has the effect that the sign switches, so one 
obtains for $f_0 < g_0$ the equality
\begin{eqnarray} \label{long_line_4}
 &&  \sum_{(M, \pi, \tau, \eta, (i,j)) \in  
\Sigma_{i_0j_0}^{f_0 g_0} \cup \Sigma_{i_0 j_0}^{g_0 f_0}} 
 \frac{sgn(\pi \pi_M (\tau \cdot \eta)^{-1})}{(m+n)!} \quad
\langle z_{f_0} * z_{g_0} |H^{(int)} \left[ w_{i_0} * w_{j_0} \right] 
\rangle \nonumber \\ &&
\prod_{l \neq i_0, j_0} 
\langle z_{\pi \pi_M (\tau \cdot \eta )^{-1} (l)} | w_l \rangle  \nonumber \\ 
&=&   sgn( \pi_{i_0 j_0}^{f_0  g_0} ) \quad
\langle z_{f_0} \otimes z_{g_0} |H^{(int)} \left[ w_{i_0} \otimes w_{j_0} 
\right] \rangle
\end{eqnarray}
in case there is a permutation $\pi_{i_0 j_0}^{f_0 g_0}$ with the demanded 
properties (\ref{properties}).

The sign can be represented as
\begin{eqnarray*}
 sgn (\pi_{i_0 j_0}^{f_0 g_0} ) &=& \langle z_1 \otimes \dots \otimes z_{m+n} |  
 c_{z_{f_0}}^\dagger c_{z_{g_0}}^\dagger c_{w_{j_0}} c_{w_{i_0}}  \left[
w_1 \otimes \dots \otimes w_{m+n}  \right] \rangle . 
\end{eqnarray*}
The equality (\ref{long_line_4}) can be rewritten as 
\begin{eqnarray*}
   && \sum_{(M, \pi, \tau, \eta, (i,j)) \in  
\Sigma_{i_0j_0}^{f_0 g_0} \cup \Sigma_{i_0 j_0}^{g_0 f_0}} 
 \frac{sgn(\pi \pi_M (\tau \cdot \eta)^{-1})}{(m+n)!} \quad
\langle z_{f_0} * z_{g_0} |H^{(int)} \left[ w_{i_0} * w_{j_0} \right] 
\rangle  \\ &&
\prod_{l \neq i_0, j_0} 
\langle z_{\pi \pi_M (\tau \cdot \eta )^{-1} (l)} | w_l \rangle   \\ 
&=&   \langle z_1 \otimes \dots \otimes z_{m+n} |
\langle z_{f_0} \otimes z_{g_0} |H^{(int)} \left[ w_{i_0} \otimes w_{j_0} 
\right] \rangle  \\ && c_{z_{f_0}}^\dagger c_{z_{g_0}}^\dagger 
c_{w_{j_0}} c_{w_{i_0}}  \left[
w_1 \otimes \dots \otimes w_{m+n}  \right] \rangle .
\end{eqnarray*}

Written in this way, the equality is correct {\em for all} 
$1 \le i_0 \le m, m+1 \le j_0 \le m+n$ and all $1 \le f_0 < g_0 \le m+n$, 
no matter whether a permutation with  the properties (\ref{properties})
exists. The sum (\ref{long_line_2}) (=(\ref{long_line_3})) turns into
\begin{eqnarray*}
 && \langle z_1 \otimes \dots \otimes z_{m+n} |
\sum_{i = 1}^m \sum_{j = m+1}^{m+n} \quad \sum_{1 \le f < g \le m+n}
\langle z_f \otimes z_g |H^{(int)} \left[ w_i \otimes w_j 
\right] \rangle  \\ && c_{z_f}^\dagger c_{z_g}^\dagger 
c_{w_j} c_{w_i}  \left[
w_1 \otimes \dots \otimes w_{m+n}  \right] \rangle .
\end{eqnarray*}
Going back to Eq. (\ref{long_line_1}), one can write the contribution of the 
electron-electron interaction to the matrix element 
\begin{displaymath}
 \langle z_1 \otimes \dots \otimes z_{m+n} | 
H_T \left[ x_1 \otimes \dots \otimes x_m \otimes y_1 \otimes \dots \otimes y_n \right]  \rangle 
\end{displaymath}
as   
\begin{eqnarray*}
 && \langle z_1 \otimes \dots \otimes z_{m+n} |
\sum_{i = 1}^m \sum_{j = 1}^{n} \quad \sum_{1 \le f < g \le m+n}
\langle z_f \otimes z_g |H^{(int)} \left[ x_i \otimes y_j 
\right] \rangle  \\ && c_{z_f}^\dagger c_{z_g}^\dagger 
c_{y_j} c_{x_i}  \left[
x_1 \otimes \dots \otimes  x_m \otimes y_1 \otimes \dots \otimes
y_{n}  \right] \rangle 
\end{eqnarray*}  
\begin{displaymath}
= \langle z_1 \otimes \dots \otimes z_{m+n} | 
H_T^{(int)} \left[ 
x_1 \otimes \dots \otimes x_m \otimes y_1 \otimes \dots \otimes y_n \right]  
\rangle, 
\end{displaymath}
where I used the definition
\begin{equation} \label{H_T^int}
 H_T^{(int)} := \sum_{\alpha, \alpha' \in R} \sum_{\beta, \beta' \in Q}
\langle \alpha' \otimes \beta' | H^{(int)} \alpha \otimes \beta \rangle \quad 
c_{\alpha'}^\dagger c_{\beta'}^\dagger c_\beta c_\alpha.
\end{equation}
$H_T^{(int)}$ is a hermitian operator and is the contribution of the 
electron-electron interaction to the total tunneling Hamiltonian.

\underline{Remark:} \newline
$H_T^{(int)}$ has the alternative representation 
\begin{displaymath} 
H_T^{(int)} = 1/2 \sum_{\alpha \in R} \sum_{\beta\in Q} \quad \sum_{\gamma, 
\delta \in R \cup Q: \gamma \neq \delta }
\langle \gamma \otimes \delta | H^{(int)} \alpha \otimes \beta \rangle \quad 
c_{\gamma}^\dagger c_{\delta}^\dagger c_\beta c_\alpha.
\end{displaymath}
To verify this, one has to show that in case $\gamma \in R$ {\em and} 
$\delta \in R$ the matrix element 
\begin{displaymath}
 \langle \gamma \otimes \delta | H^{(int)} \alpha \otimes \beta \rangle 
\end{displaymath}
vanishes. One writes this as 
\begin{displaymath}
 \langle \left[ (H^{(int)} - H^R) + H^R \right]\gamma \otimes \delta |  
\alpha \otimes \beta \rangle
\end{displaymath}
and uses that $H^R (\gamma \otimes \delta) \in \overline { V (Q^\bot)} $ according 
to the complementarity-condition (\ref{orthogonality}). Because $\beta \in Q$, 
the contribution of $H^R$ vanishes. A close look and another application of the 
complementarity yield that also the contribution of $H^{(int)}
- H^R$ vanishes.

\subsection{Contribution of the potentials to $H_T$}
Next, the contribution of $T_{m,n}^{(Q)}$ to the matrix element of Eq. 
(\ref{long_line_1}) is determined. For this I assume $m \ge 1$ and write in the same 
way as in the previous subsection (Eq. \ref{definition_of_w-s}):
\begin{displaymath}
 x_{\tau (1)} * \dots * x_{\tau (m)} *  y_{\eta (1)} * \dots * y_{\eta (n)} =
w_{(\tau \cdot \eta) (1)} * \dots * w_{(\tau \cdot \eta) (1)}.
\end{displaymath}
In analogy to the previous subsection, one can write for any $1 \le i \le m$ and 
any sequence $v_1, \dots , v_{m+n}$ of one-electron wave functions 
\begin{eqnarray*}
 V^{Q}(r_i) \left[ v_1 * \dots * v_{m+n} \right] =
S_{\kappa_i} \left[ \left( V^{Q}(r) v_{\kappa_i (1)}   \right)
* v_{\kappa_i (1)} * \dots * v_{\kappa_i (m+n)}    \right] ,  
\end{eqnarray*}
where the permutation $\kappa_i \in S_{m+n}$, which is defined by 
the conditions
\begin{itemize}
 \item $\kappa_i (1) = i  $,
\item $\kappa_i (2), \dots , \kappa_i (m+n)$ is an increasing (arbitrary) list of 
the remaining numbers $\neq i$, 
\end{itemize}
was used. One obtains the contribution
\begin{eqnarray*}
 \sum_{M, \pi, \tau, \eta} \sum_{i = 1}^m \frac{sgn(\pi \pi_M 
(\tau \cdot \eta)^{-1} ) }    {(m+n)!} && \\
\langle z_{ \pi \pi_M (\tau \cdot \eta)^{-1} (\tau \cdot \eta) \kappa_i (1) } |
V^{Q}(r) w_{(\tau \cdot \eta) \kappa_i (1)} \rangle \quad && \\
\prod_{l=2}^{m+n} \langle z_{ \pi \pi_M (\tau \cdot \eta)^{-1} (\tau \cdot \eta) 
\kappa_i (l) } | w_{(\tau \cdot \eta) \kappa_i (l) } \rangle 
\end{eqnarray*}
of the potential $V^{Q}$ to the matrix element (\ref{long_line_1}). For every
$1 \le i_0 \le m  $ and every $1 \le f_0 \le m+n$ I define the set 
$\Sigma_{i_0}^{f_0}$ as the set of all tuples 
\begin{displaymath}
 ( M, \pi, \tau, \eta, i ) 
\end{displaymath}
with the property that 
\begin{eqnarray*}
 (\tau \cdot \eta) \kappa_i (1) &=& i_0, \\
\pi \pi_M  \kappa_i (1) &=& f_0. 
\end{eqnarray*}
The set of all tuples over which the sum goes is the disjoint union of the sets
$\Sigma_{i_0}^{f_0}$. Thus, the sum can be rewritten as
\begin{eqnarray*}
\sum_{i_0, f_0} 
\sum_{(M, \pi, \tau, \eta, i) \in \Sigma_{i_0}^{f_0}} \frac{sgn(\pi \pi_M 
(\tau \cdot \eta)^{-1} ) }    {(m+n)!} && \\
\langle z_{f_0} |
V^{Q}(r) w_{i_0} \rangle \quad && \\
\prod_{l\neq i_0} \langle z_{ \pi \pi_M (\tau \cdot \eta)^{-1} (l)  } | 
w_l  \rangle. 
\end{eqnarray*}

For fixed $i_0, f_0$, there is at most one permutation ``$\pi_{i_0}^{f_0} $'' with 
the properties 
\begin{itemize}
 \item $ \prod_{l\neq i_0} \langle z_{ \pi_{i_0}^{f_0} (l)  } | w_l  \rangle \neq 0 $
\item $ \pi_{i_0}^{f_0} (i_0) = f_0$.
\end{itemize}
{\em If} there is such a permutation, then the number of all tuples 
$( M, \pi, \tau, \eta, i ) \in \Sigma_{i_o}^{f_0} $ with the property that 
\begin{displaymath}
 \pi \pi_M (\tau \cdot \eta)^{-1} =  \pi_{i_0}^{f_0}
\end{displaymath}
is $(m+n)! $. The inner sum has in this case the value 
\begin{eqnarray*}
 && sgn( \pi_{i_0}^{f_0} ) \quad \langle z_{f_0} | V^{Q}(r) w_{i_0} \rangle  \\
&=&  \langle z_1 \otimes \dots \otimes z_{m+n} | 
\quad \langle z_{f_0} | V^{Q} (r) w_{i_0} \rangle 
c_{z_{f_0} }^\dagger c_{w_{i_0}} \left( w_1 \otimes \dots \otimes w_{m+n} 
\right) \rangle .
\end{eqnarray*}

The representation of the inner sum by the second line is correct even in the case 
that there is {\em no} permutation with the properties of $\pi_{i_0}^{f_0}$.
Thus, the contribution of the potential $V^{Q}$ to the matrix element of Eq. 
(\ref{long_line_1}), 
\begin{displaymath}
 \langle z_1 \otimes \dots \otimes z_{m+n} | 
H_T \left[ x_1 \otimes \dots \otimes x_m \otimes y_1 \otimes \dots \otimes y_n 
\right]  \rangle, 
\end{displaymath}
is
\begin{eqnarray*}
 \langle z_1 \otimes \dots \otimes z_{m+n} | && \\ 
 \sum_{i=1}^m \sum_{f=1}^{m+n} \langle z_{f} | V^{Q} (r) x_{i} \rangle \quad
c_{z_{f} }^\dagger c_{x_{i}}  && \\ \left[ x_1 \otimes \dots \otimes x_m \otimes
y_1 \otimes \dots \otimes y_{n} 
\right] \rangle \quad =  &&\\
\langle z_1 \otimes \dots \otimes z_{m+n} | && \\ 
 \sum_{\alpha \in R} \quad \sum_{\beta \in R \cup Q} \langle \beta | V^{Q} (r) 
\alpha \rangle \quad
c_{\beta }^\dagger c_{\alpha}  && \\ \left[ x_1 \otimes \dots \otimes x_m \otimes
y_1 \otimes \dots \otimes y_{n} 
\right] \rangle .
\end{eqnarray*}

The sum of the contributions of the potentials $V^{Q}$ and $V^{R} $ to the 
tunneling Hamiltonian is 
\begin{eqnarray} \label{H_T^V}
 H_T^{(V)} &=& \sum_{ S \in \left\lbrace R, Q \right\rbrace } \quad 
\sum_{\alpha, \alpha' \in S} \quad \langle \alpha | V^{\bar{S}} (r) \alpha' \rangle
\quad c_\alpha^\dagger c_{\alpha'}  \nonumber \\
& + & \sum_{S \in \left\lbrace R, Q \right\rbrace } \quad \sum_{\alpha \in S}
\sum_{\beta \in \bar{S}} \quad \frac{1}{2} \langle \alpha | V(r) \beta \rangle
\quad c_\alpha^\dagger c_{\beta}.  
\end{eqnarray}
I used the letter ``$S$`` to denote the two sub-systems; ``$\bar{S} $'' is the 
the complementary sub-system of $S$; $V(r)$ is the sum of the two potentials, 
$V = V^{R} + V^{Q}$.

\underline{Remark:} \newline
$H_T^{(V)}$ has the alternative representation 
\begin{displaymath}
H_T^{(V)} =  \sum_{ S \in \left\lbrace R, Q \right\rbrace } \quad
\sum_{\alpha \in S} \quad \sum_{\beta \in R \cup Q} \langle \beta | 
V^{\bar{S}} (r) 
\alpha \rangle \quad
c_{\beta }^\dagger c_{\alpha}. 
\end{displaymath}
To see the identity, one can use the assumed complementarity (\ref{orthogonality}) 
and write for $\alpha \in R, \beta \in Q$:
\begin{eqnarray*}
 \langle \alpha | V^{R} \beta \rangle &=& 
\langle (H^R - c \Delta) \alpha | \beta \rangle \quad = \quad   
\langle - c \Delta \alpha | \beta \rangle  \\
&=& \langle \alpha | -c \Delta \beta \rangle \quad =  \quad 
\langle \alpha | - ( H^Q - V^{Q} ) \beta \rangle \\
&=& \langle \beta | V^{Q} \alpha \rangle ^\ast, 
\end{eqnarray*}
where I used the abbreviation $c:= - \frac{\hbar^2}{2m}$.

\subsection{Expression for $H_T$ in terms of creation- and annihilation operators:
Summary}
The tunneling Hamiltonian, defined by Eq. (\ref{definition_of_H_T}), 
is the sum of the contributions of the electron-electron interaction, 
Eq. (\ref{H_T^int}), and that of the potentials, Eq. (\ref{H_T^V}):
\begin{eqnarray*}
 H_T &=& \sum_{\alpha, \alpha' \in R} \sum_{\beta, \beta' \in Q}
\langle \alpha' \otimes \beta' | H^{(int)} \alpha \otimes \beta \rangle \quad 
c_{\alpha'}^\dagger c_{\beta'}^\dagger c_\beta c_\alpha  \\
&+& \sum_{ S \in \left\lbrace R, Q \right\rbrace } \quad 
\sum_{\alpha, \alpha' \in S} \quad \langle \alpha | V^{\bar{S}} (r) \alpha' \rangle
\quad c_\alpha^\dagger c_{\alpha'}  \nonumber \\
& + & \sum_{S \in \left\lbrace R, Q \right\rbrace } \quad \sum_{\alpha \in S}
\sum_{\beta \in \bar{S}} \quad \frac{1}{2} \langle \alpha | V(r) \beta \rangle
\quad c_\alpha^\dagger c_{\beta}.
\end{eqnarray*}

\subsection{Example}
I would like to illustrate a possible effect of the additional contributions
$H_T^{(int)}$, Eq. (\ref{H_T^int}), to the tunneling Hamiltonian. If I assume 
that $\alpha, \alpha' \in R, \beta, \beta' \in Q$ with 
\begin{eqnarray*}
 \alpha (r,\sigma) &=& \psi_{lk} (r) \delta_\downarrow (\sigma), \\
 \alpha' (r,\sigma) &=& \psi_{l'k'} (r) \delta_\uparrow (\sigma), \\ 
\beta (r, \sigma) &=& \varphi_{\uparrow} (r) \delta_{\uparrow} (\sigma), \\
\beta' (r, \sigma) &=& \varphi_{\downarrow} (r) \delta_{\downarrow} (\sigma), 
\end{eqnarray*}
then the corresponding contribution to the tunneling Hamiltonian reads
\begin{equation} \label{example}
 -e^2 \int\int dr dr' \frac{\varphi_\downarrow (r)^\ast \psi_{lk\downarrow} (r) 
\varphi_\uparrow (r') \psi_{l'k'\uparrow} (r')^\ast}{|r - r'|}  \quad
c_{\alpha'}^\dagger c_{\beta'}^\dagger c_\beta c_\alpha.  
\end{equation}
The term annihilates an electron with spin ``down'' in the leads as well as an
electron on the quantum dot with spin ``up''. At the same time, it creates 
an electron with spin ``down'' on the quantum dot and an electron with spin 
``up'' in the leads. The term represents a spin-flip process and simultaneous 
scattering of electrons in the leads. The coefficient which is obtained in case
$l' = l$ can be expected to be larger compared to the case $l' \neq l$
($l$ is the lead index).

\section{Outlook}
By taking the reduced density matrix and applying the projection operator technique
\cite{Blum}, it might be possible to approach the transport problem with two 
leads at different chemical potentials even when terms as in the example 
of the last subsection are included. A complete (diagrammatic) analysis of the kernels
as in Ref. \cite{Schoeller97} would be desirable, but not necessary for perturbation
theory of lowest order in the tunneling Hamiltonian. The terms in $H_T$ of the form 
\begin{displaymath}
 \langle \alpha | V(r) \beta \rangle \quad c_\alpha^\dagger c_{\beta}, \quad \alpha
\in S, \beta \in \bar{S},
\end{displaymath}
give rise to energy conserving single electron tunneling seen by perturbation 
theory of lowest order in the tunnel coupling, e.g. Ref. \cite{Schoeller97}. 
Analogously, one might expect that from the  terms of the form of the example 
(\ref{example}) one obtains transition rates
\begin{displaymath}
 \Gamma_{ll'} ( \uparrow \rightarrow \downarrow )
\end{displaymath}
whose value is sensitive to the question whether there are electrons and holes with
appropriate spins in the leads, such that the spin-flip- and scattering 
process obeys the energy conservation. This might have interesting consequences 
for the current across the quantum dot, especially if $l \neq l'$.

\section{Conclusions}
I considered the Hamiltonian of an electronic system which consists of two 
weakly interacting sub-systems. The starting point was the many-electron Hamiltonian 
obtained from the Born-Oppenheimer approximation. Moreover, it was assumed that 
the wave functions of the states of the composite system can be constructed 
by the use of the wave functions of the states of the sub-systems in a natural 
way (antisymmetric product). The assumption of weak interaction became manifest in 
a ``complementarity condition'' (\ref{orthogonality}). It turns out that the 
total Hamiltonian has a natural additive separation into three contributions:
For each of the sub-systems one obtains a contribution which is essentially given
by the Hamiltonian of the corresponding sub-system.

The contribution which expresses the interaction is called tunneling Hamiltonian.
An analysis of it showed that it can be represented in a natural way in terms of 
annihilation- and creation operators. The contributions to the tunneling Hamiltonian
are classified according to their origin: One kind of terms is caused by the 
electron-electron interaction, while another kind of terms is due to the interaction
of the positive nuclei and the electrons.

The considerations are very general: The inner electronic structure of the sub-
systems was not relevant. Even if the electron-electron interaction can be taken 
into account in the sub-systems by an effective noninteracting Hamiltonian, 
it is still expected that in the tunneling Hamiltonian the terms caused by the 
electron-electron interaction appear.

In my mind and in the text I used the picture that the systems are distant 
and separated by vacuum. As an example, the conditions for the derivations are 
approximately fulfilled also in the case of impurities in a metal. The sub-systems 
are then the impurity and the metal. The overlap of the one-electron wave functions 
can be expected to be small if the level generated by the impurity is localized.
Apart from the transport across a quantum dot, one possible way to see whether the 
spin-flip scattering processes contained in 
the tunneling Hamiltonian yield sensible results would be to apply them to the 
problem of a resistivity minimum in metals doped with magnetic impurities
\cite{de Haas}, explained first in Ref. \cite{Kondo} by spin-flip scattering 
processes. The formal similarity of the problem with the transport across quantum 
dots led to the prediction of the zero bias resonance known as Kondo resonance
\cite{Wilhelm02, Goldhaber98, Ralph94}. Spin-flip scattering processes with even
a similar {\em formulation} as in the example (\ref{example}) have been taken into 
account for example in Ref. \cite{Bruus}. The coefficients of the terms are 
different. Nevertheless, the qualitative behaviour of the resulting current 
might be expected to be similar.

\section{Appendix}

\subsection{Separation of $H$ into one- and two-particle operators}
The total Hamiltonian, Eq. (\ref{Hamiltonian}), with general potential $V(r)$
has the natural additive separation 
\begin{displaymath}
 H = H^{(int)} + H_V, 
\end{displaymath}
where $H^{(int)}$ is the contribution of the electron-electron interaction, 
Eq. (\ref{interaction}), and where by ``$H_V$'' the operator
\begin{displaymath}
 H_V = - \frac{\hbar^2}{2m} \Delta  + \sum_{i}  V( r_i )
\end{displaymath}
is denoted. The separation is obtained from a classification into 
one-particle operators ($H_V$) and two-particle operators 
($H^{(int)} $), Ref. \cite{Bruus}. I tried to express or verify this
classification also by the use of the antisymmetric product.

By methods very similar to the ones applied in the main part of the text 
one obtains for the action of $H_V$ and $H^{(int)}$ on any Slater determinant
of one-electron wave functions $\psi_1, \dots , \psi_N$:
\begin{eqnarray*}
 H^{(int)} \left[ \psi_1 \otimes \dots \otimes \psi_N \right] &=& 
\sum_{1 \le f < g \le N} sgn (\kappa_{fg}) \left[ 
H^{(int)} ( \psi_f \otimes \psi_g )   \right] \otimes   \\ && \quad \quad \quad
\quad \quad \psi_{\kappa_{fg}(3)}
\otimes \dots \otimes \psi_{\kappa_{fg} (N)}, \\ 
H_V   \left[ \psi_1 \otimes \dots \otimes \psi_N \right] &=& 
\sum_{1 \le f \le N} sgn(\kappa_f) \left[ H_V \psi_f \right] \otimes  \\ &&
\quad \quad \quad \quad
\psi_{\kappa_f (2)} \otimes \dots \otimes \psi_{\kappa_f (N)}.
\end{eqnarray*}
The permutations $\kappa_f, \kappa_{fg} \in S_N$ are defined in the same way as 
in the main text,i.e., $\kappa_f (1) = f,  \kappa_{fg} (1) = f, 
\kappa_{fg}(2) = g$, the rest is irrelevant.

For the action of the operators on a vector space $V(R)$ generated by an arbitrary
orthonormal system $R$ of one-electron wave functions one obtains
in terms of creation- and annihilation operators:
\begin{eqnarray*}
 H^{(int)} \psi &=& \sum_{\alpha \in R} (H_V \alpha) \otimes (c_\alpha \psi ), \\
H_V \psi   &=& \frac{1}{2} \sum_{\alpha \neq \beta } 
\left[ H^{(int)} ( \alpha \otimes \beta )  \right] \otimes 
\left[ c_\beta c_\alpha \psi \right]  
\end{eqnarray*}
for arbitrary $\psi \in V(R)$.

If $p_{V(R)}$ is the orthonormal projection onto $V(R)$, then the operator
$p_{V(R)} H = p_{V(R)} ( H^{(int)} + H_V )$ has the representation
\begin{eqnarray} \label{free electrons}
 p_{V(R)} H &=&  \frac{1}{4} \sum_{\alpha, \beta, \gamma, \delta \in R} \quad 
\langle \gamma \otimes \delta | H^{(int)} (\alpha \otimes \beta ) \rangle  \quad
c_\gamma^\dagger c_\delta^\dagger c_\beta c_\alpha \nonumber  \\  && + 
\sum_{\alpha, \gamma \in R} \quad \langle \gamma | H_V \alpha \rangle  \quad 
c_\gamma^\dagger c_\alpha .     
\end{eqnarray}
If the orthonormal system $R$ is chosen as a system of eigenfunctions of 
the one-electron operator $H_V$, then the second line of Eq. 
(\ref{free electrons}) turns into 
\begin{displaymath}
 \sum_{\alpha \in R} \quad \lambda_\alpha  \quad 
c_\alpha^\dagger c_\alpha, 
\end{displaymath}
where $\lambda_\alpha$ is the eigenvalue of the eigenfunction $\alpha$ of 
$H_V$. The latter operator is diagonal in the Slater determinants of the one-electron 
wave functions in $R$.

The potential $V(r)$ is caused by the nuclei in the crystal. If the crystal 
is perfectly regular, then the potential $V(r)$ can be written as a sum of a periodic
contribution and a rather slowly varying non-periodic contribution: For every 
$r \in \mathbb{R}^3$ one can imagine a ball with centre $r$ and positive and fixed
radius. The contribution of the nuclei {\em within} the ball is periodic in the 
lattice. The contribution of the nuclei outside this ball is slowly 
varying and has larger negative values in the centre compared to its values 
closer the fringe. Therefore, the eigenfunctions of $H_V$ might be expected to 
produce an electron density which is not periodic, but rather concentrated in the 
centre of the crystal. This is, at least in a vague sense, inconsistent with the 
fact that, {\em if} the lattice of the nuclei is perfectly periodic, then the 
density of the positive charges is also periodic and {\em not} concentrated 
in the centre of the crystal. With this reasoning I would assume that it might 
be an {\em inadequate} ansatz to try to find or get close to eigenfunctions of the 
total Hamiltonian by using Slater determinants of eigenfunctions of $H_V$. 
Even if these Slater determinants would turn out to be good approximate 
eigenfunctions of $H$, it would seem unlikely that an eigenfunction of the final 
Hamiltonian (including also the nuclei) can be constructed by the use of the
Slater determinants.

\subsection{Hybridization}
Alternatively, I wrote for the action of the Hamiltonian, Eq. (\ref{Hamiltonian}),
on an $N$-electron wave function $\psi$ with arbitrary $N \ge 2$:
\begin{eqnarray*}
 H \psi &=& \sum_{1 \le i<j \le N} \left\lbrace 
\frac{e^2}{|r_i - r_j|} + \frac{1}{N-1} \left[ V(r_i) + V(r_j) + c  
( \Delta_i + \Delta_j ) \right] \right\rbrace \psi  \\
&=:& \sum_{1 \le i<j \le N} H_{ij}^{(N)} \psi .
 \end{eqnarray*}
With the definition 
\begin{displaymath}
 H^{(N)} := \frac{e^2}{|r_1 - r_2|} + \frac{1}{N-1} \left[ V(r_1) + V(r_2) + c  
 \Delta  \right]
\end{displaymath}
(an operator acting on two-electron wave functions only) one can represent the
action of $H$ on an arbitrary Slater determinant of $N$ one-electron wave functions
as
\begin{eqnarray*}
  H \left[ \psi_1 \otimes \dots \otimes \psi_N \right] &=& 
\sum_{1 \le f < g \le N} sgn (\kappa_{fg}) \left[ 
H^{(N)} ( \psi_f \otimes \psi_g )   \right] \otimes   \\ && \quad \quad \quad
\quad \quad \psi_{\kappa_{fg}(3)}
\otimes \dots \otimes \psi_{\kappa_{fg} (N)} .
\end{eqnarray*}

If $R$ is an orthonormal system of one-electron wave functions, then one obtains for 
$\psi \in V_N (R)$: 
\begin{displaymath}
 H \psi = \frac{1}{2} \sum_{\alpha, \beta \in R} 
\left[ H^{(N)} (\alpha \otimes \beta ) \right] \otimes \left[ 
c_\beta c_\alpha \psi \right]
\end{displaymath}
and 
\begin{displaymath}
 p_{V(R)} H \psi = \frac{1}{4} \sum_{\alpha, \beta, \gamma, \delta \in R} \quad 
\langle \gamma \otimes \delta | H^{(N)} (\alpha \otimes \beta ) \rangle  \quad
c_\gamma^\dagger c_\delta^\dagger c_\beta c_\alpha \quad \psi 
\end{displaymath}

The representation of the action of $H$ in terms of the operators $H^{(N)}$
seems appealing, compared to the separation into $H^{(int)}$ and $H_V$. 
Since any antisymmetric $N$-electron wave function can be written as an 
(infinite) linear combination of Slater determinants, the description of 
the action of $H$ on the space of the $N$-electron wave functions is 
{\em equivalent} to the description of the action of the operator $H^{(N)}$
on two-electron wave functions. The representations of $H$ obtained in the 
appendix are independent of the shape of the potential $V(r)$.

\section*{Acknowledgments}
I thank the DFG for financial support within the framework of the GRK 1570. 
Moreover, I thank Piotr Chudzinski for his comment on a previous version of this 
text.

\end{document}